\documentstyle[aaspp]{article}
\begin{document}
\lefthead{Peletier \& Balcells}
\righthead{Colors and Ages of Disks and Bulges}

\def\rpcomm#1{{\bf COMMENT by RP: #1} \message{#1}}
\def\square{\vrule height 4.5pt width 4pt depth -0.5pt}
\def\threesquares{\square~\square~\square\ }
\def\remark#1{{\threesquares\tt#1~\threesquares}}
\def\ie{{\it i.e.}}
\def\eg{{\it e.g.}}
\def\deg{\ifmmode^\circ\else$^\circ$\fi}
\def\Msun{M_\odot}
\def\etal{{\it et al.~}}
\def\deg{$^{\rm o~}$}
\def\bck{\hskip-0.35em}
\def\min#1{\ifmmode  {^{\prime}}                           
            \else    {$^{\prime}$}\fi
            \ifcat,#1{\bck}\else\null\fi\ #1}
\def\deg{\ifmmode {^{\rm o}}              
         \else {$^{\rm o}$}\fi}
\def\sec{\ifmmode {^{\prime\prime}~}       
         \else {$^{\prime\prime}~$}\fi}
\def\me{$^{-1}$}              

\onecolumn
\title{Ages of galaxy bulges and disks from optical and near-infrared colors}

\author{R. F. Peletier \and M. Balcells}

\affil{Kapteyn Astronomical Institute,
	Postbus 800, 9700~AV~~Groningen, Netherlands}

\begin{abstract}
We compare optical and near-infrared colors of disks and bulges 
in a diameter-limited sample of inclined, bright, nearby, early-type spirals.  
Color profiles along wedge apertures at 15\deg\ from the 
major axis and on the minor axis on the side of the 
galaxy opposite to the dust lane
are used to assign nominal colors for the inner disks (at 2 scale length) 
and for the bulges ($\sim$ 0.5 $r_{eff}$), respectively.  
We estimate  that the effects of dust reddening and the cross-talk
between the colors of the two components is negligible.
We find that color differences (bulge -- disk) are very small:
$\Delta(U-R)=0.126 \pm 0.165$, $\Delta(R-K)=0.078 \pm 0.165$.  
Disks tend to be bluer by an amount three times smaller
than that reported by Bothun \& Gregg \markcite{B84} (1990) for S0's.  
Color variations from galaxy to galaxy are much larger than 
color differences between disk and bulge in each galaxy.
Probably, the underlying old population of disks and bulges is
much 
more similar than the population paradigm would lead us to believe.  
Implied age differences, assuming identical metallicities, 
are less than 30\%.
\end{abstract}


\section{Introduction}

How different are the colors of bulges and disks of spiral galaxies? 
According to the Population paradigm, 
disks of spirals, containing Population I stars,
are bluer and younger than bulges, the latter made up of Population II stars.
Also, the widespread belief that bulges are metal rich (eg. Rich 
\markcite{R88} 1988)
supports the notion that bulges are redder than disks.
This notion needs to be reexamined in the light of the following:
(1) We find that bulges of early-type spirals are not metal rich; 
their mean metallicities are rarely above solar 
(Balcells \& Peletier \markcite{BP94} 
1994, hereafter BP94). Also the metallicity
of our Bulge nowadays is estimated to be just above solar (McWilliam
\& Rich \markcite{MR94} 1994).
(2) The inner parts of disks are often very dusty (Valentijn \markcite{V90} 1990,  
Peletier \etal \markcite{PVMFKB95} 1995).
Therefore, integrated colors (and colors derived from ellipse fitting, 
see section \ref{Data}) trace reddening as much as population characteristics. 
(3) Colors have become useful age diagnostic tools
with the realization by Frogel \markcite{F85} (1985) that 
the combination of optical and near-infrared (NIR) color indices
allows to partially disentangle age and metallicity.
(4) Integrated colors include contributions from bright,
localized, blue star forming regions.  
The colors of the underlying disk contain
information on the age of the older components of the disk 
population.  Excluding regions of ongoing star formation 
and regions with significant reddening
from the color determinations was not possible with aperture photometry,
but can now be done using two-dimensional digital photometry 
and modern data processing techniques.

The recent availability of large-format two-dimensional near-infrared
detectors allows the determination of NIR color profiles
without the limitations of studies based on NIR aperture photometry data.
In this paper we analyze $UBRIJK$ surface photometry of
a sample of early-to-intermediate type, inclined disk galaxies.
This work is part of a larger study of optical and NIR colors
of disks and bulges (Peletier \& Balcells, in preparation).
Here we highlight that the colors of 
bulges and disks are very similar.  
Indeed, 
contrary to what would be expected if bulges are indeed redder than disks,
bulges do not appear as distinct morphological components
in color index maps derived from optical and NIR images.
We find long dust lanes to be the main morphological component 
of the color maps. 
A few galaxies show additional features such as  nuclear spots of redder color
with a size smaller than the bulge,
and progressive  bluing in the outer parts of the galaxy.
Color variations from galaxy to galaxy are much larger
than color variations within each galaxy.

To put this result in a more quantitative footing, 
we work with color profiles of disk and bulge determined
in a way which minimizes the effects of dust reddening. 
Our results, from accurate near-infrared array measurements,
show that color differences between bulges and disks are very small.
Age determinations derived from optical and NIR color-color diagrams
with the use of simple population models 
indicate that the bulk of the (inner) disk and bulge stars 
are essentially coeval.  At most, disk stars are 2-3 Gyr 
younger than bulge stars.  
We disagree with the conclusion of Bothun \& Gregg \markcite{BG90} (1990)
that disks are bluer than bulges by $\Delta(B-H) \sim 0.4$.

\section{The data}
\label{Data}

\begin{figure}
\label{Plate}
\plotone{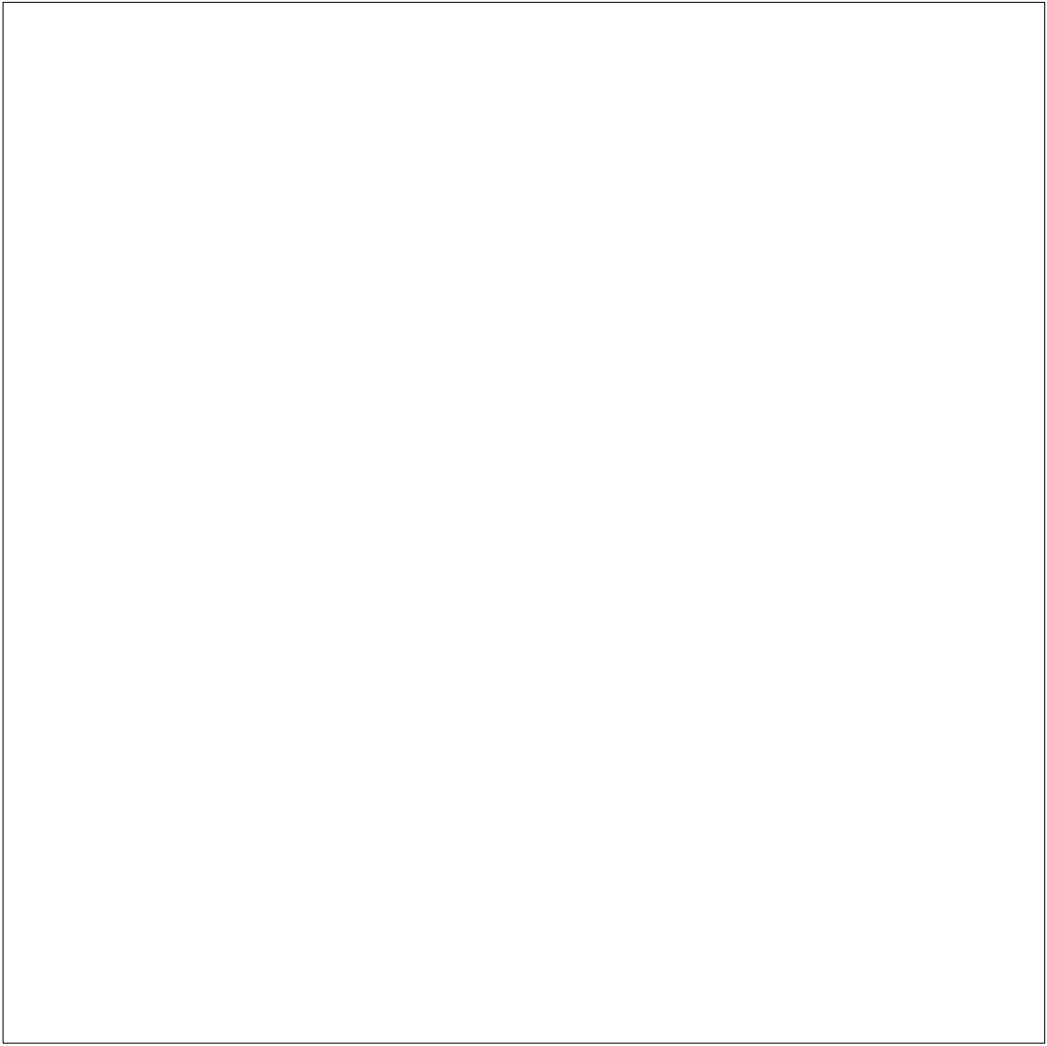}
\caption{
	Real-color plate of the 30 galaxies. In the RGB representation
	we have used the $U$ images for the blue colors, the $R$ images
	for green, and the $K$-band images for the red colors. Owing to the
	photometric calibration, the color scheme is uniform for all 
	galaxies, i.e. bluer galaxies appear bluer on the plate.}
\end{figure}

Our sample consists of inclined galaxies (i $\geq$ 50\deg), of  
galaxy type S0 to Sbc. The sample comprises the 30 galaxies of Table 1
of BP94, except for the two that were 
outside the declination range of UKIRT. The optical colors of the bulges
have been analyzed in that paper. In short,
the optical data consist of U,B,R and I surface photometry, obtained
with a 400$\times$590 EEV CCD chip
at the PF camera of the 2.5m INT telescope at La Palma.
The pixelsize was 0.549'', and the effective seeing on the images lies between
1.2'' and 1.6''. The images were taken under photometric conditions, and
flatfielding was accurate to about 0.2\%. 
The colors were corrected for Galactic extinction (BP94).
We obtained near-infrared images for this sample
at the United Kingdom Infrared Telescope (UKIRT) using
IRCAM3 (Puxley \& Aspin \markcite{PA94} 1994), an infrared camera
equipped with a 256 $\times$ 256 InSb array.
The camera gives a pixel size of 0.291 arcsec and a field of 75 arcsec.
Cosmetically, the array is very clean, with less than 1\%  bad pixels.
For every object we took images in 10 positions, 
of several readouts each, 
making up a total integration time of 100\,s per
position. The object was moved around on the chip on 6 of these
exposures, while 4 consisted of blank sky, about 10 arcmin away from the galaxy.
The frames were flatfielded using median sky flat fields, and 
mosaics were made aligning the individual frames. The effective seeing
on the final frames was between 0.8'' and 1.0''. The frames
here were also taken under photometric conditions, 
with errors for individual stars of $\sim$0.03 mag. Comparison
with aperture photometry shows that the maximum zero point
errors are 0.1 mag in $J$ and $K$. All galaxies were observed in $K$. 
In addition, 20 galaxies were also observed in the $J$-band. 
In Fig.~\ref{Plate} we show 'real-color' URK-images  for this sample.
To make the images we first calibrated the frames in each band photometrically,
and then aligned the three images using the positions of stars.

\begin{table}
\begin{center}
\caption{
	\label{ColTab}
	Colors of bulges and disks
	}
\begin{tabular}{lcccccccc}
\hline
\hline
Galaxy & (B-R)$_B$ & (B-R)$_D$ & (U-R)$_B$ & (U-R)$_D$ & 
(R-K)$_B$ & (R-K)$_D$ & (J-K)$_B$ & (J-K)$_D$ \\
\hline

NGC   5326   &   1.444  &    1.405   &   1.946   &   1.845  &    2.604  &     2.56  &    0.874  &    0.911 \\
NGC   5362   &   1.144  &     0.97   &    1.46   &   1.012  &    2.414  &    2.274  &     --    &     --   \\
NGC   5389   &   1.439  &    1.434   &   1.741   &   1.772  &    2.678  &    2.662  &    0.874  &    0.921 \\
NGC   5422   &   1.462  &    1.376   &   1.901   &   1.804  &    2.704  &    2.709  &    0.867  &    0.913 \\
NGC   5443   &   1.521  &    1.354   &   2.057   &   1.776  &    2.461  &    2.485  &    0.931  &    0.925 \\
NGC   5475   &   1.474  &    1.506   &   1.876   &   1.885  &    2.498  &    2.474  &    0.867  &    0.883 \\
NGC   5577   &     1.3  &    1.061   &   1.235   &   0.821  &    2.539  &    2.171  &    0.835  &    0.589 \\
NGC   5587   &   1.471  &    1.423   &   1.914   &   1.777  &    2.564  &    2.602  &     --    &     --   \\
NGC   5675   &   1.472  &    1.329   &     --    &    --    &    2.58   &    2.458  &     --    &     --   \\
IC    1029   &   1.402  &     1.25   &   1.808   &   1.614  &    2.492  &     2.52  &     --    &     --   \\
NGC   5689   &   1.437  &    1.482   &   1.861   &   1.896  &    2.705  &    2.638  &    0.902  &    0.918 \\
NGC   5707   &    1.53  &    1.328   &   1.874   &   1.629  &    2.711  &    2.588  &     --    &     --   \\
NGC   5719   &    1.61  &    1.402   &   2.059   &   1.759  &    2.933  &    2.435  &     --    &     --   \\
NGC   5746   &   1.595  &     1.65   &   2.162   &   2.214  &    2.821  &    2.851  &    0.982  &    0.979 \\
NGC   5838   &   1.488  &    1.435   &   2.039   &   1.871  &    2.725  &    2.679  &    0.866  &    0.838 \\
NGC   5866   &   1.451  &    1.457   &   1.841   &   1.682  &    2.609  &     2.44  &    0.863  &     0.94 \\
NGC   5854   &   1.392  &    1.442   &   1.742   &    1.79  &     2.48  &    2.495  &    0.922  &    0.781 \\
NGC   5879   &   1.271  &    1.314   &   1.279   &   1.293  &    2.404  &    2.476  &    0.779  &    0.833 \\
NGC   5908   &   1.441  &    1.547   &   1.765   &   1.908  &    2.595  &    2.984  &     --    &     --   \\
NGC   5965   &   1.408  &    1.444   &   1.978   &   2.038  &    2.667  &    2.731  &    0.901  &    0.962 \\
NGC   5987   &    1.49  &    1.477   &   1.962   &   1.886  &    2.966  &    2.667  &     1.18  &    1.074 \\
NGC   6010   &   1.435  &    1.504   &   1.805   &   1.759  &    2.535  &     2.66  &    0.954  &    1.046 \\
NGC   6368   &   1.723  &    1.607   &   2.031   &    1.91  &    3.121  &    2.976  &    0.896  &    0.929 \\
NGC   6504   &   1.789  &    1.746   &   2.118   &   1.721  &    2.633  &    2.381  &     --    &     --   \\
NGC   6757   &   1.434  &     1.38   &   1.672   &    1.46  &    2.707  &     2.76  &     --    &     --   \\
NGC   7311   &   1.523  &    1.348   &   1.963   &   1.479  &    2.827  &    2.726  &     --    &     --   \\
NGC   7332   &   1.353  &    1.413   &   1.802   &   1.753  &    2.393  &    2.165  &    0.829  &    0.658 \\
NGC   7457   &   1.287  &    1.325   &   1.675   &   1.618  &    2.403  &    2.172  &    0.862  &    0.829 \\
NGC   7537   &   1.285  &    1.242   &   1.267   &   1.138  &    2.593  &    2.379  &    0.875  &    0.866 \\
NGC   7711   &   1.405  &    1.464   &   2.022   &   2.083  &    2.653  &    2.551  &    0.901  &    0.838 \\
\hline
\end{tabular}
\end{center}
\end{table}

Color profiles for bulges and disks were obtained by 
measuring surface brightness profiles on each band
along wedge-shaped apertures.
For the bulge profiles, wedges were 22.5\deg\ wide, 
centered on the disk semi-minor axes as determined by elliptical fits
to the galaxy's image.  
In galaxies with prominent dust lanes,
only the semi-minor axis away from the dust lane was used
for subsequent analysis.
Otherwise, we took the mean of the two profiles.
Details on the method are given in Balcells \& Peletier \markcite{BP94} (1994). 
The disk surface brightness profiles were measured
along 10\deg-wide wedge apertures 
centered 15\deg\ away from the disk major axis,
as determined in the $K$-band image.
We chose these apertures so as to avoid measuring over
the prominent dust lanes near the major axis of inclined galaxies.  
We thus measure the light of the disk on the far side of the galaxy
which is not extincted by the dust on the near side of the disk. 
The color profiles could be slightly affected 
by any vertical disk color gradients for inclinations above 80\deg.

We chose the wedge-aperture method of determining color profiles over
the more common method 
of integrating azimuthally along best-fit ellipses
(Terndrup \etal  \markcite{T94} 1994; Peletier \etal \markcite{PVMF94} 
1994; de Jong \markcite{dJ95} 1995).  
For galaxies at high inclinations, such as those in our sample,
the two methods give very different results.
The difference is higher for galaxies with redder colors,
a sign that the dust is responsible for the observed differences. 
The effects of dust in the wedge color profiles
can be estimated from the amplitude of the disk color gradients
in various bands.  
For our sample, dust effects appear to be small 
for galaxy types up to Sab and for some Sb's.

\section{Colors and color differences}
\label{Colors}

\begin{figure}
\plotone{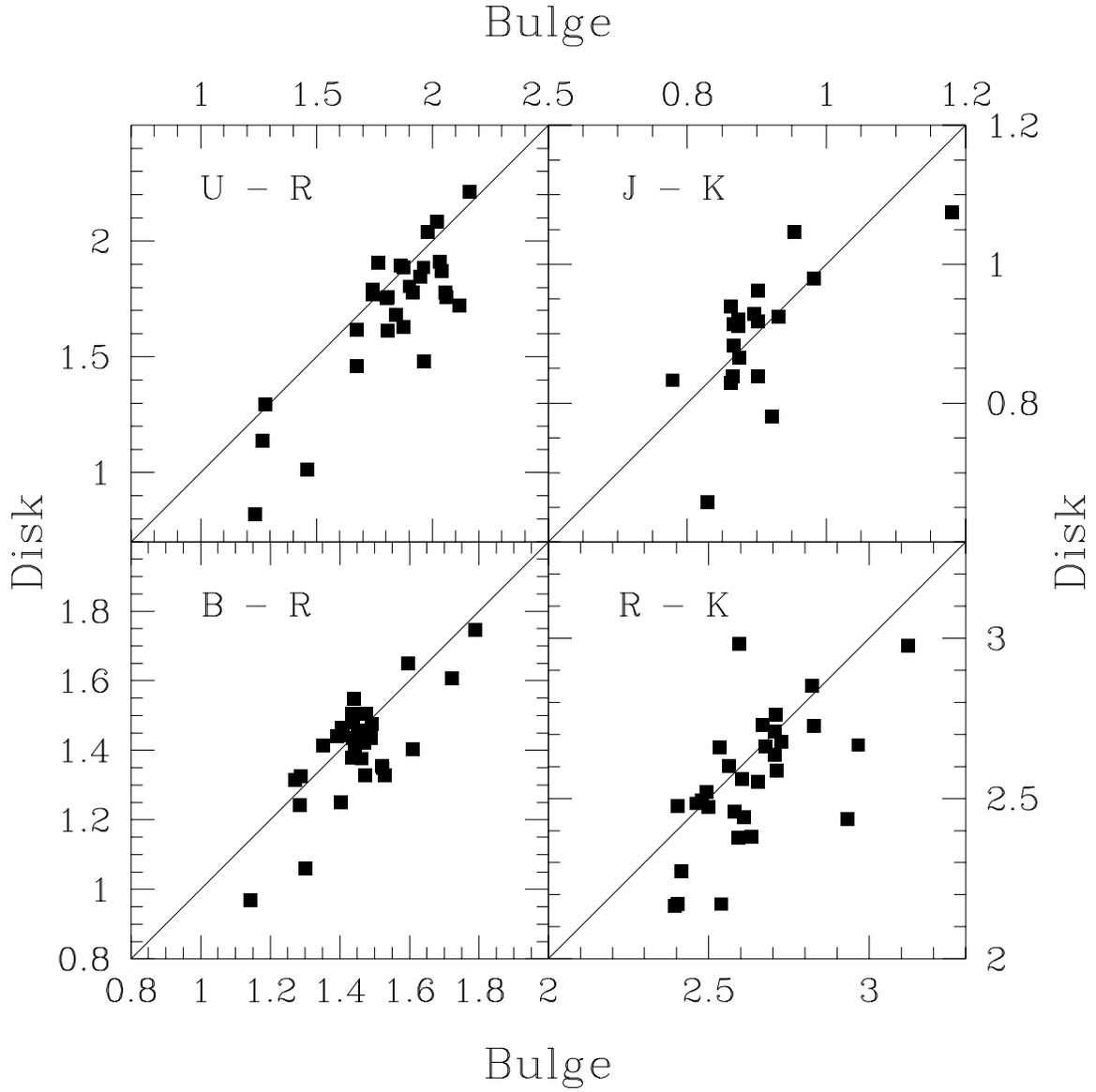}
\caption{
\label{BulDisCol}
	Disk colors as a function of bulge colors, disk
	colors taken at 2 major axis scale lengths, 
	bulge colors at r$_e$/2 or 5'', whichever is larger.}
\end{figure}

Colors in bulges and disks become bluer radially outward 
(de Jong, \markcite{dJ95} 1995; \markcite{BP94} BP94).  
Gradients are small enough however that 
we assign representative values for the color of each component.  
For bulges, we take the color at 0.5$\times R_{eff}$ 
or at 5~arcsec, whichever is larger.  
For disks we use the colors at 2 major axis $K$-band scale lengths.  
{These colors are tabulated in Table~1.}
We find that disk colors, while somewhat bluer,
are very similar to bulge colors for all the galaxies
(see Fig.~\ref{BulDisCol}).
In Fig.~\ref{BulDisCol},
the diagonal line indicates the locus where both colors are equal.
The average differences between disk and bulge colors are
0.126 $\pm$ 0.165 for $U$ -- $R$, 0.045 $\pm$ 0.097 for
$B$ -- $R$, 0.078 $\pm$ 0.165 for $R$ -- $K$ 
and 0.016 $\pm$ 0.087 for $J$ -- $K$ 
($J$ -- $K$ colors are only available for 20 of the 30 galaxies).
Larger deviations tend to occur at the blue end of each 
distribution.  This region of the diagrams is occupied
by low-luminosity, late-type galaxies. Note that the average difference
between bulge and disk color is three times
smaller than the one reported by Bothun \& Gregg \markcite{BG90} (1990) for S0's.
Since color gradients in these bulges and inner disks are small
(BP94, Peletier \& Balcells, in preparation) this result does not
critically depend upon the place where the colors are measured.
Terndrup \etal (1994) also show that $R-K$ colors of bulges and
disks are not very different. The average $R-K$ difference between
their bulge and disk color is 0.21 mag, slightly larger than what we find.
The interpretation of this number however is much more difficult
for their sample, since it is unclear how much their colours are affected
by extinction.

\begin{figure}
\plotone{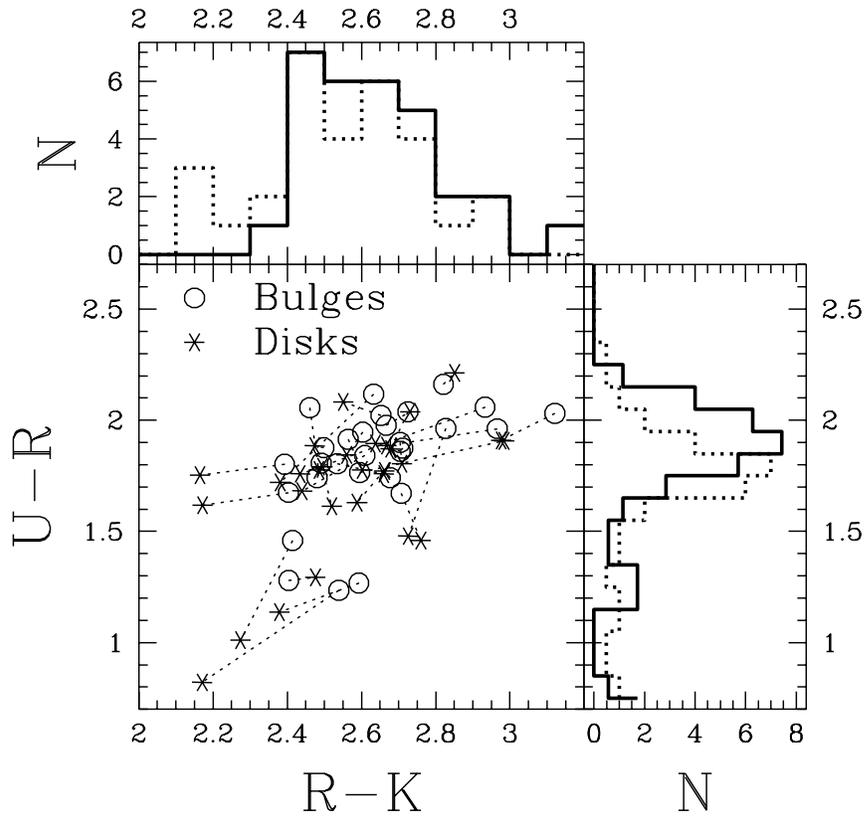}
\label{URK}
\caption{
Color-color diagram indicating bulges (circles)
and disks (asterisks). Dotted lines connect 
the colors of each disk with those of the corresponding bulge.
The histograms to the top give the distributions of $R-K$ 
for bulges (solid lines) and disks (dotted lines).  
The histograms to the right give the distributions of $U-R$ 
for bulges (solid lines) and disks (dotted lines).  
}
\end{figure}

We now put the information contained in Fig.~\ref{BulDisCol}
as a blue vs. infrared color-color diagram.  
Figure~\ref{URK} shows the $U$--$R$ vs. $R$--$K$ diagram 
for bulges (circles) and disks (asterisks).  
Dashed lines connect the bulge and disk for each galaxy.
On the sides, histograms are given for the $U-R$ and $R-K$
colors of the galaxies. Figures~\ref{BulDisCol} and \ref{URK} show that
the colors of the stellar populations
of bulges and inner disks are very similar. This is true for the
bluest galaxies, as well as for the other, redder, galaxies.
The distributions of $U-R$ and $R-K$ are not identical though,
as shown in the histograms to the right and to the top 
of Fig.~\ref{URK} (bulges: solid lines.  Disks:  dotted lines).  
The $U-R$ distributions are clearly different, in a KS sense. 
The $R-K$ distributions are undistinguishable in a KS test 
except for a blue tail in the color distribution of disks. 

\section{Discussion}

\begin{figure}
\plotone{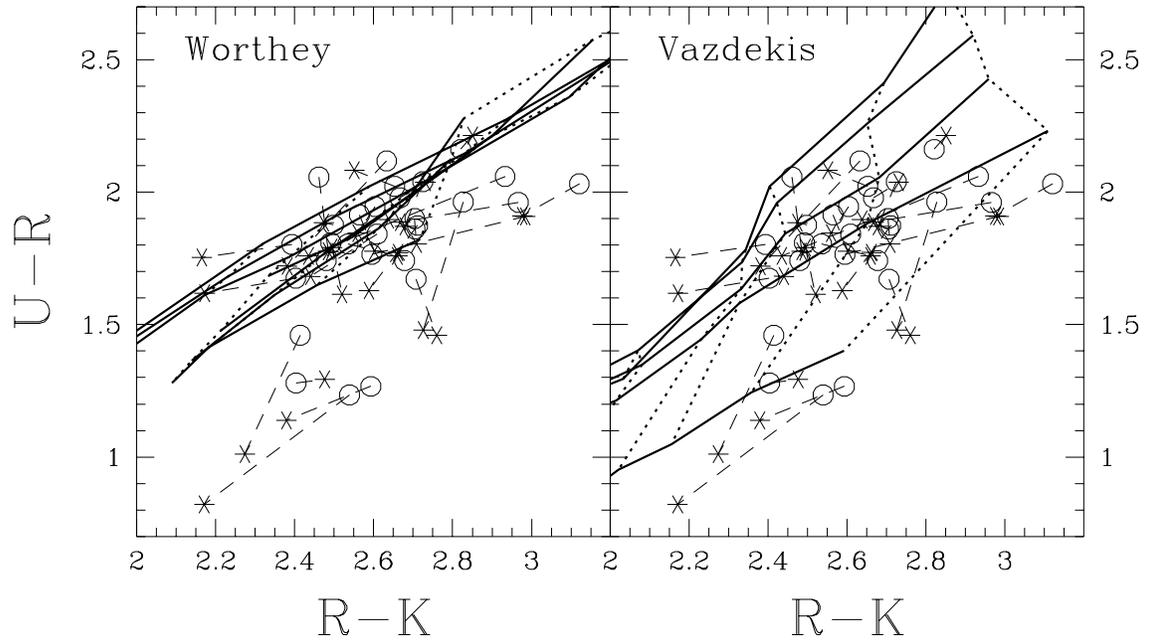}
\label{TwoGrid}
\caption{
Stellar population models for single age,
single metallicity stellar population models by Worthey (1994)
and Vazdekis \etal (1995) 
superimposed on top of the data of Fig.~\ref{URK}.
Drawn lines are lines of constant age, dotted lines of
constant metallicity.}
\end{figure}

Can we use the color similarity to put limits on the age differences
between bulges and disks? 
In Fig.~\ref{TwoGrid} we show the color-color data of Fig.~\ref{URK}
together with single 
age-metallicity stellar population models by Worthey \markcite{W90} (1994) and
Vazdekis \etal \markcite{V95} (1995). Along the solid lines, metallicity varies, 
while age varies along the dotted lines. It is clear that there are 
large differences between the color predictions of both models. 
Color and metallicity are partially disentangled 
in the models of Vazdekis \etal and degenerate 
in those of Worthey.
Such differences may be due to difficulties inherent to 
the modeling of $U$ and $K$ data. 
In the $U$-band, few stars are available to calibrate 
the models with. In the $K$-band, model predictions 
are affected by uncertainties 
in the treatment of the later stages of stellar evolution,
such as the AGB (see Charlot \etal 1996 for a discussion).
Vazdekis \etal \markcite{V95} (1995) are not the first to claim that age
and metallicity are separable using this diagram:
this notion already appears in the models
of Bothun \etal \markcite{B84} (1984), Peletier \etal \markcite{PVJ90} 
(1990), and Bressan \etal \markcite{BCF94} (1994).

The agreement among the latter authors prompts us to 
use the Vazdekis \etal\ models to analyze our data.  
Color differences are so small however, that
a separation of age and metallicity will not be attempted. 
Instead, we use the models to see what age difference
is implied by the color differences in the case of 
constant metallicity, 
and what metallicity difference is implied by the color differences
in the case of constant age.  
In Table~2 and 3 we give the difference in metallicity
while keeping age constant, and the difference in age, while metallicity
is left constant. The models used are the two mentioned above
at Z=0.02 and age=12 Gyr, plus the models of Rabin \markcite{R80} 
(1980) as given by
O'Connell \markcite{O86} (1986). The latter models have been modified slightly using
Worthey's values, to convert $U-V$, $B-V$ and $V-K$ to resp. 
$U-R$, $B-R$ and $R-K$. Table~2 shows again that there are large discrepancies
between the 3 models. These show up especially in $R-K$.
On the average however, the models agree that one needs
a difference of appr. 0.10 in log~Z, or 0.11 in log~t 
between bulge and disk. At an age of 10 Gyr, this
corresponds to a difference of 3 Gyr. It is however much more likely that
the difference in color is caused by both metallicity and age.
In this case the age difference is smaller.
We stress that these results apply to the inner parts of disks
only. Photographs of spirals often suggest that the outer parts
are bluer, and perhaps the perception many people have about the
blueness of disks really applies to the regions at large radii.

\begin{table}
\begin{center}
\caption{Color - metallicity/age conversions}
\begin{tabular}{ccccc}
\\
\\
\hline
Parameter & & Vazdekis & Worthey & Rabin \\
\hline
$\Delta(B-R)$/$\Delta(log~Z)$ & = & 0.37 & 0.48 & 0.54  \\
$\Delta(U-R)$/$\Delta(log~Z)$ & = & 0.79 & 1.28 & 1.46  \\
$\Delta(R-K)$/$\Delta(log~Z)$ & = & 0.62 & 1.51 & 0.61  \\
$\Delta(J-K)$/$\Delta(log~Z)$ & = & 0.17 & 0.29 & --    \\
$\Delta(B-R)$/$\Delta(log~t)$ & = & 0.47 & 0.34 & 0.65 \\
$\Delta(U-R)$/$\Delta(log~t)$ & = & 1.09 & 0.79 & 1.29 \\
$\Delta(R-K)$/$\Delta(log~t)$ & = & 0.04 & 0.72 & 0.42 \\
$\Delta(J-K)$/$\Delta(log~t)$ & = & -0.06 & 0.16 & -- \\
\hline
\hline
\end{tabular}
\end{center}
\end{table}
\begin{table}
\begin{center}
\caption{Age and metallicity differences between bulges and disks}
\begin{tabular}{cccccccc}
\\
\\
\hline
Parameter & Color & \multicolumn{2}{c}{Vazdekis} & 
\multicolumn{2}{c}{Worthey} & \multicolumn{2}{c}{Rabin} \\
 & & Average & Spread & Average & Spread & Average & Spread \\
\hline
$\Delta(log~Z)$ & B -- R & --0.12 & 0.26 & --0.09 & 0.20 & --0.08 & 0.18 \\
 & U -- R & --0.16 & 0.21 & --0.10 & 0.13 & --0.09 & 0.11 \\
 & R -- K & --0.13 & 0.26 & --0.05 & 0.11 & --0.13 & 0.27 \\
 & J -- K & --0.10 & 0.52 & --0.05 & 0.30 & -- & -- \\
$\Delta(log~t)$ & B -- R & --0.10 & 0.28 & --0.13 & 0.28 & --0.07 & 0.15 \\
 & U -- R & --0.12 & 0.15 & --0.16 & 0.21 & --0.10 & 0.13 \\
 & R -- K & --2.11 & 4.46 & --0.11 & 0.23 & --0.19 & 0.39 \\
 & J -- K & 0.28 & 1.50   & --0.10 & 0.56 & -- & -- \\
\hline
\end{tabular}
\end{center}
\end{table}


Our result is different from that of Bothun \& Gregg \markcite{BG90} 
(1990), who find
that in the $B-H$ vs. $J-K$ diagram disks of S0's are systematically displaced
with respect to bulges. For a given $J-K$ the $B-H$ of a disk
is on the average 0.4 mag bluer than a bulge. 
Here, we don't find any systematic offset between bulges and disks,
as they see for example in their Fig.~7.
To explain the discrepancy
we can only point to the fact that their infrared (JHK) measurements
are aperture measurements, where the aperture centers are not centered
on the galaxy nucleus. Making these measurements, and comparing them
to CCD data is extremely complicated, since results depend for
example on the response map of the aperture itself, the wings of the 
aperture, etc. (see e.g. Peletier \etal \markcite{PVJ90} 
1990). An error of 0.5 mag
seems plausible to us. 
For the S0's in our sample
we find that there is as much, or as little,
star formation in the bulge as in the disk. 

Our data put constraints on the Eggen \etal\ \markcite{ELS62} (1962) 
model of galaxy formation by monolithic collapse with 
progressive enrichment of the galaxy. Our ages make it
unlikely that there has been
a gap in time between the formation of the bulge and that of the inner 
disk. The formation of bulge and disk must have been a continuous process.
This is the case for example in the continuous infall models
of Gunn \markcite{G82} (1982), where the age of the stars is determined by the 
free-fall collapse time of the infalling gas. Since this time-scale 
is almost equal for bulge and inner disk, no stellar population differences
will be expected.

The similarities in color between bulges and disks can be believed
to stem naturally from models in which the bulge forms 
from instabilities in the disk 
(Pfenniger \& Norman \markcite{PN90} 1990; Combes et al. \markcite{C90} 
1990; Pfenniger \markcite{P93} 1993).
This is indeed the case in the absence of starburst processes; 
starbursts imply discontinuous changes 
which break the age and metallicity link between parent
and daughter population.  
Starbursts are likely to be necessary in order to build up the 
central space densities of bulges from disk material;
they may be unavoidable if the disk contains any gas.    
Thus, our result does not necessarily support instability-driven 
bulge formation scenarios.

Our result, that an important part of the disk is as old as the bulge,
also means that, at high redshift, a considerable amount of gas must already
have been transformed from gas into stars. 
We predict that the occurrence of "naked bulges"
surrounded by large amounts of gas is unlikely at any redshift.
The statistics of galaxies 
associated with damped Lyman-alpha systems,
still in low numbers, 
reveal the predominance of disk morphologies (Wolfe \etal \markcite{W94} 1994).

How do color gradients in bulge and disk affect this result? Measured 
gradients in both bulges and disks are generally small (BP94,
Peletier \& Balcells 1995, in preparation), so age or metallicity 
only differ by a small amount if another position in the bulge
or disk is chosen. 
Fisher \etal \markcite{FFI95} (1995) report high values of color and line-strength gradients
in S0's.  
The difference with our data is that their systems are extreme edge-on 
galaxies. It appears that vertical gradients in bulges are large, contrary
to radial gradients. We find that the color gradients in the inner
effective radius of the bulges are small, and comparable in size to
those in elliptical galaxies (BP94, Peletier \& Balcells, in preparation), so
the discrepancy with Fisher \etal\ might be due to the fact that,
owing to the almost perfect edge-on aspect,
their measurements go further out.

The main conclusion of this work is that, consistently 
from 4 independent colors, we find that inner disks are only slightly younger
than bulges. The difference in age in general lies somewhere between
0 and 3 Gyr.

\acknowledgments
We thank Alexandre Vazdekis for 
making his stellar population models available before publication and
Massimo Stiavelli for help with the observations. We thank K. Kuijken,
A. Renzini and D. Fisher for very helpful comments on the manuscript.

\end{document}